\documentclass{IEEEtran}

\usepackage[english]{babel}

\usepackage{algorithm}
\usepackage{algorithmicx}
\usepackage{algpseudocode}
\usepackage{multirow}
\usepackage{amsmath}
\usepackage{amsopn}
\usepackage{amssymb}

\usepackage{caption}

\usepackage{cite}

\usepackage{CJK}

\ifCLASSOPTIONcompsoc
    \usepackage[tight,normalsize,sf,SF]{subfigure}
\else
    \usepackage[tight,footnotesize]{subfigure}
\fi

\ifCLASSINFOpdf
  \usepackage[pdftex]{graphicx}
  \usepackage{epstopdf}
  \graphicspath{{./pic/}}
  \DeclareGraphicsExtensions{.pdf,.jpeg,.png}
\else
  \usepackage[dvips]{graphicx}
  \DeclareGraphicsExtensions{.eps}
\fi

\begin{document}
\title{FDRC: Flow-Driven Rule Caching Optimization in Software Defined Networking}

\author{\IEEEauthorblockN{He Li\IEEEauthorrefmark{1},
                          Song Guo\IEEEauthorrefmark{1},
                          Chentao Wu\IEEEauthorrefmark{2},
                          and Jie Li\IEEEauthorrefmark{2},
                          }\\
\IEEEauthorblockA{
\IEEEauthorrefmark{1}The University of Aizu\\
Aizuwakamatsu, Fukushima, Japan \\
}

\IEEEauthorblockA{\IEEEauthorrefmark{2}Shanghai Key Laboratory of Scalable Computing and Systems, Department of Computer Science \& Engineering, Shanghai Jiao Tong University, Shanghai, China 200240\\}
\IEEEauthorblockA{\{d8141105, sguo\}@u-aizu.ac.jp, {\{wuct, lijie\}@cs.sjtu.edu.cn} \\
                  }
}

\maketitle

\begin{abstract}
With the sharp growth of cloud services and their possible combinations, the scale of data center network traffic has an inevitable explosive increasing in recent years. \textit{Software defined network} (SDN) provides a scalable and flexible structure to simplify network traffic management. It has been shown that \textit{Ternary Content Addressable Memory} (TCAM) management plays an important role on the performance of SDN. However, previous literatures, in point of view on rule placement strategies, are still insufficient to provide high scalability for processing large flow sets with a limited TCAM size. So caching is a brand new method for TCAM management which can provide better performance than rule placement. In this paper, we propose FDRC, an efficient flow-driven rule caching algorithm to optimize the cache replacement in SDN-based networks. Different from the previous packet-driven caching algorithm, FDRC is characterized by trying to deal with the challenges of limited cache size constraint and unpredictable flows. In particular, we design a caching algorithm with low-complexity to achieve high cache hit ratio by prefetching and special replacement strategy for predictable and unpredictable flows, respectively. By conducting extensive simulations, we demonstrate that our proposed caching algorithm significantly outperforms FIFO and \textit{least recently used} (LRU) algorithms under various network settings.

\end{abstract}

\section{Introduction}
\subsection{Rule space structure}

As one of the most significant technologies for large-scale data center network, \textit{Software Defined Network (SDN)} demonstrates great potentiality on management, scalability and other features. SDN-enabled switches, managed by a logically centralized controller, support fine-grained and flow-level controls of data center networks. Such controls are desirable with flexible policies under programmable configuration and visible flow management \cite{Koponen2010,Curtis2011,Levin2012,Monsanto2013}. Typically, the flow-based control is implemented by installing simple packet-processing rules in the underlying switches. These rules can match the head of packet-header fields in the network flows, and perform some actions, such as forwarding, modifying, or sending to controller for further processing. For each flow, combining multiple policies can provide a flexible, fine-grained and dynamical control. However, with the increasing number of flows in data center networks, the flow-based control leads to a combinatorial explosion in terms of all number of rules per switch.

In previous SDN switches, \textit{Ternary Content Addressable Memory} (TCAM) is a typical memory to store rules, which can compare an incoming packet to the patterns in all of rules at a line rate simultaneously \cite{McKeown2008}. However, TCAM is not a cost-effective way to provide high performance. First, compared to the ordinary RAM, TCAM needs approximate 400 times monetary cost and consumes 100 times power. Second, even in high-end commodity switches, due to the limited size of TCAM, the space cannot contain a large number of various rules. Furthermore, the updating speed of rules is slow in TCAM, which supplies around 50 rule-table updates per second. This is a major restriction for adopting policies to support flow-based control in large-scale networks \cite{Qazi2013}.

Rule placement optimization is an existing method to improve the processing capacity of data center networks \cite{Liu2010, Meiners2011, Meiners2012}. The mechanism of this method is that by getting the information on the whole network, after some analysis on all existing flows, a proper placement strategy can be found to improve the flow processing capacity. However, these optimizing strategies are usually static with a limited number of flows. Furthermore, when the status of flows changes, such as flow destination movement, traffic variation, etc., updating the rules in all switches is unaffordable.

Unlike rule placement optimization which regards rule space as a limited resource, rule caching strategies efficiently use space to store the rules and cache the most frequent rules in TCAM as caching \cite{Kanizo2013, Moshref2013, Kang2013}. Therefore, all rules can be handled in the network via replacement policies and the performance can be enhanced in terms of high hit ratio. Compared to the rule placement optimization methods, rule caching is a better approach to enable the flow-based control to provide both high performance and scalability, especially in a large-scale data center network.

Ordinary caching algorithms, such as \textit{Least Recently Used} (LRU), are not appropriate to the flow-based control in large-scale data center networks. One reason is that these algorithms are based on single rule replacement for each switch, while proper control policies should be based on a global view of multiple rules in all switches. Furthermore, the fact that flow traffic is predictable and correlated with time should be exploited. Through acquiring and analyzing the history flow information, an SDN controller can easily get the significant information of applications, i.e., the prediction of flow traffic distribution, such that higher performance can be archived.

To address this problem on rule caching, in this paper, we model the optimization problem and design a caching algorithm based on the prefetching and replacement strategies. This algorithm achieves high hit ratio by replacing rules with the flow forwarding paths and replacing them integrally with different replacement strategies for predictable and unpredictable flows.

The main contributions of this paper are summarized as follows,
\begin{itemize}
\item First, we study a caching optimization problem for flow-based control in the data center networks. This problem is challenging because of the flow variability and the constrained cache space in SDN switches.

\item Second, we propose an efficient heuristic algorithm to solve the caching optimization problem. Our basic idea is to design two different replacement strategies for predictable and unpredictable flows.

\item Finally, extensive simulations are conducted and the results show that the proposed algorithm can significantly increase the hit ratio and thus improve the network performance.
\end{itemize}

The rest of this paper are organized as follows. 
Our network model and system design are introduced in Section \ref{sec:bm}. Section \ref{sec:pss} presents our algorithm design. Section \ref{sec:s} gives simulation results. Finally, Section \ref{sec:c} concludes this paper.

\section{System Model}
\label{sec:bm}

In this section, we first discuss the rule caching and flow-based control in SDN-based network. Then, we state the main problem in the rule caching. For better understanding, we use Table \ref{tab:notation} to show the meanings of major notations.
\begin{table}[h]
\centering
\caption{Notations}
\label{tab:notation}
\begin{tabular}{ll}
 Notation  &  Description \\
\hline $f_i$  & Flow $i$  \\
$F$  & Flows in the network \\
 $f_i$  & Flow $i$  \\
 $R$ & Rules in the network \\
 $r_i$  &   Rule $i$ of flow $i$ \\
 $S$  & Switches in the network \\
 $S_i$ & Switches in the forwarding path of flow $i$ \\
 $s_j$  & Switch $j$ \\
 $B_j$  & Cache size of switch $j$ \\
$f_i(t)$ & Network traffic density function of flow $i$ at time $t$\\
$X_{ij}$ & Whether switch $j$  caches rule of flow $i$ \\
$h_i(t)$ & Cache Hits of rule $i$ at time $t$ \\
$H_i(t)$ & Cache hits of rule $i$ from time 0 to $t$ \\
$F_i(t)$ & Traffic of flow $i$ from time 0 to $t$ \\
$C_i(t)$ & Cache hit ratio of flow $i$ from time 0 to $t$ \\
$C(t)$ & Cache hit ratio of entire network \\
$T(f_i, t)$ & Time to the next coming packet of $f_i$ at time $t$ \\
$T_{max}$ & Maximum waiting time for the next packet coming\\
$T_{next}$ & Possible maximum time to the next packet coming \\

\end{tabular}

\end{table}
\subsection{Flows and Rule Caching}


Unlike traditional networks, SDN considers network flows as the basic units and control methodologies are based on the flows in typical. However, the rule updating in SDN switches is related to the network packets. A rule updating takes place during the processing of new network packets. When a new packet is checked and no matching field with the entries of flow tables in the switch, the packet will be sent to the controller for further processing. In general, after the processing of a new packet, the related rules are updated in the switch. This strategy is considered as a FIFO replacement.

FIFO is not a good replacement algorithm because that some rules for processing rare flows can stay in the TCAM for a long period. In a general network, many flows only have several packets in a short period of time. By using FIFO replacement policy, the rules of these flows cost too much TCAM space with a low cache hit ratio.

\begin{figure}[h]
\centering
\includegraphics[width=\linewidth]{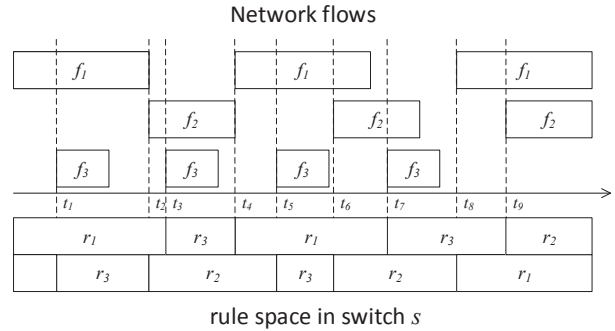}
\caption{Caching rules for network flows}
\label{fig:rc_prostate}
\end{figure}

\textit{Least Recently Used} (LRU) is an advanced caching algorithm used in many fields. Exiting solutions also use LRU for SDN caching replacement. However, it is not an appropriate caching algorithm in SDN since LRU is still a packet-driven algorithm. To illustrate the LRU in flow-based control SDN, we use an example shown in Figure \ref{fig:rc_prostate}, where three flows,  $f_1$, $f_2$ and $f_3$ are processed using rules $r_1$, $r_2$ and $r_3$, respectively. Suppose each switch cache can store two rules. For simplifying the problem, we consider the traffic the traffic of three flows regularly distributes in the time-domain cannot be interrupted (e.g., during the period from $t_2$ to $t_4$). Initially, since there is no rule cached in the TCAM space, there are two cache misses when flow $f_1$ and $f_3$ come. After that, when $f2$ comes to the switch , the algorithm replaces $r_3$ to $r_2$ with a cache miss because $r_3$ is the least recently used rule. However, since the LRU algorithm does not know that $f_3$ will come back soon, $r_3$ needs to be re-deployed to replace $r_1$. Finally, from the result of LRU replacement, the cache hit ratio in this example is 0.

Another problem is that most of flows in a data center network are relevant with more than one switches. While existing caching algorithm is packet-driven, the replacement only occurs in a single switch. When some other switches forward a same packet, controller need to process this packet again. In an SDN structure, all switches are managed by a centralized controller and the controller can also get the traffic information of each flow in the network. As a result, a flow-driven caching is more appropriate than a packet-driven algorithm with SDN. That's why we propose a novel rule caching algorithm to state the flow-driven caching problem.

\subsection{Rule Caching Problem}
\begin{figure}[h]
\centering
\includegraphics[width=\linewidth]{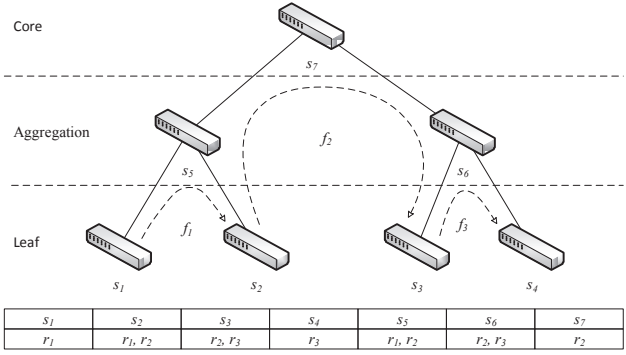}
\caption{The network flows and rule caching in a typical data center topology}
\label{fig:rule_cache_structure}
\end{figure}

As shown as Figure \ref{fig:rule_cache_structure}, we consider a data center network consisting of a set $S = \{s_1, s_2, ..., s_{m}\}$ of switches which includes leaf switches, aggregation switches and core switches. For any switch $s_j$ in $S$, it maintains a TCAM-based flow table which can cache at most $B_j$ forwarding rules.

We consider set $F = \{f_1, f_2, ... , f_{n}\}$ of network flows, with an associated set $R=\{r_{1},r_{2},...,r_{n}\}$ of forwarding rules among SDN switches. Let $S_i$ denote the set of switches in the forwarding path of $f_i$, i.e, any switch in $S_i$ maintains $r_i$. For example, $S_1 = \{s_1, s_2, s_5\}$. 

In this paper, we investigate a rule caching problem under our system model by addressing the following two challenges. First, rule capacity of each switch is limited, and the rules required by a network flow may be forwarded by multiple switches, or even cannot be cached by the SDN-based network. We define a variable $X_{ij}$ to denote the rule caching as follows.

\begin{equation}
X_{ij} = \begin{cases}
 1 \ r_i \text{ is cached in } s_j  \\
 0 \ r_i \text{ is not cached in } s_j
\end{cases}
\end{equation}

\noindent The cache capacity constraint at each switch can be represented by:

\begin{equation}
\sum_{i=1}^{n}X_{ij} \leq B_j \text{.}
\end{equation}

\noindent Letting $f_i(t)$ be the traffic density of flow $f_i$, we can define the cache hits at time instance as follows.

\begin{equation}
\label{eq:hit_t}
h_i(t) = f_i(t)\sum_{s_j\in S_i} X_{ij}
\end{equation}

\noindent Therefore, the cache hits $H_i (t)$ from time 0 to $t$ can be expressed as:

\begin{equation}
\label{eq:hit_zero_t}
H_i(t) = \int_0^t h_i(t)dt = \int_0^t f_i(t)\sum_{s_j\in S_i} X_{ij}dt\text{.}
\end{equation}

\noindent Finally using $F_i(t)$  to denote the overall traffic from time 0 to $t$, i.e.,

\begin{equation}
\label{eq:flow_zero_t}
F_i(t) = \int_0^t f_i(t) dt\text{,}
\end{equation}

\noindent we can get the hit ratio $C_i(t)$ from 0 to $t$ as follows.

\begin{equation}
\label{eq:hit_ratio_t}
C_i(t) = \frac{H_i(t)}{F_i(t)}= \frac{\int_0^t f_i(t)\sum_{s_j\in S_i} X_{ij}dt}{\int_0^t f_i(t)dt}
\end{equation}

\noindent Similarly, we define $C(t)$ to denote the total hit ratio shown in (\ref{eq:total_hit}).

\begin{equation}
\label{eq:total_hit}
C(t) = \frac{\sum_{i = 1}^{n}H_i(t)}{\sum_{i = 1}^{n}F_i(t)} = \frac{\sum_{i = 1}^{n}\int_0^t f_i(t)\sum_{s_j\in S_i} X_{ij}dt}{\sum_{i = 1}^{n}\int_0^t f_i(t) dt}
\end{equation}

\textbf{The problem of rule caching in SDN-based network}: given a software defined network, a set of network flows with rules and a period, the rule caching problem attempts to find a part of flows and put rules of these flows to the cache of each SDN switch of the network to maximum the accumulated number of  cache hits in this period.

\

\section{Algorithm}
\label{sec:pss}

\begin{figure*}[t]
  \begin{minipage}[t]{0.33\linewidth}
  \centering
\subfigure[flow $f_1(t)$]{
\label{fig:flow_predict}
\includegraphics[width=1\linewidth]{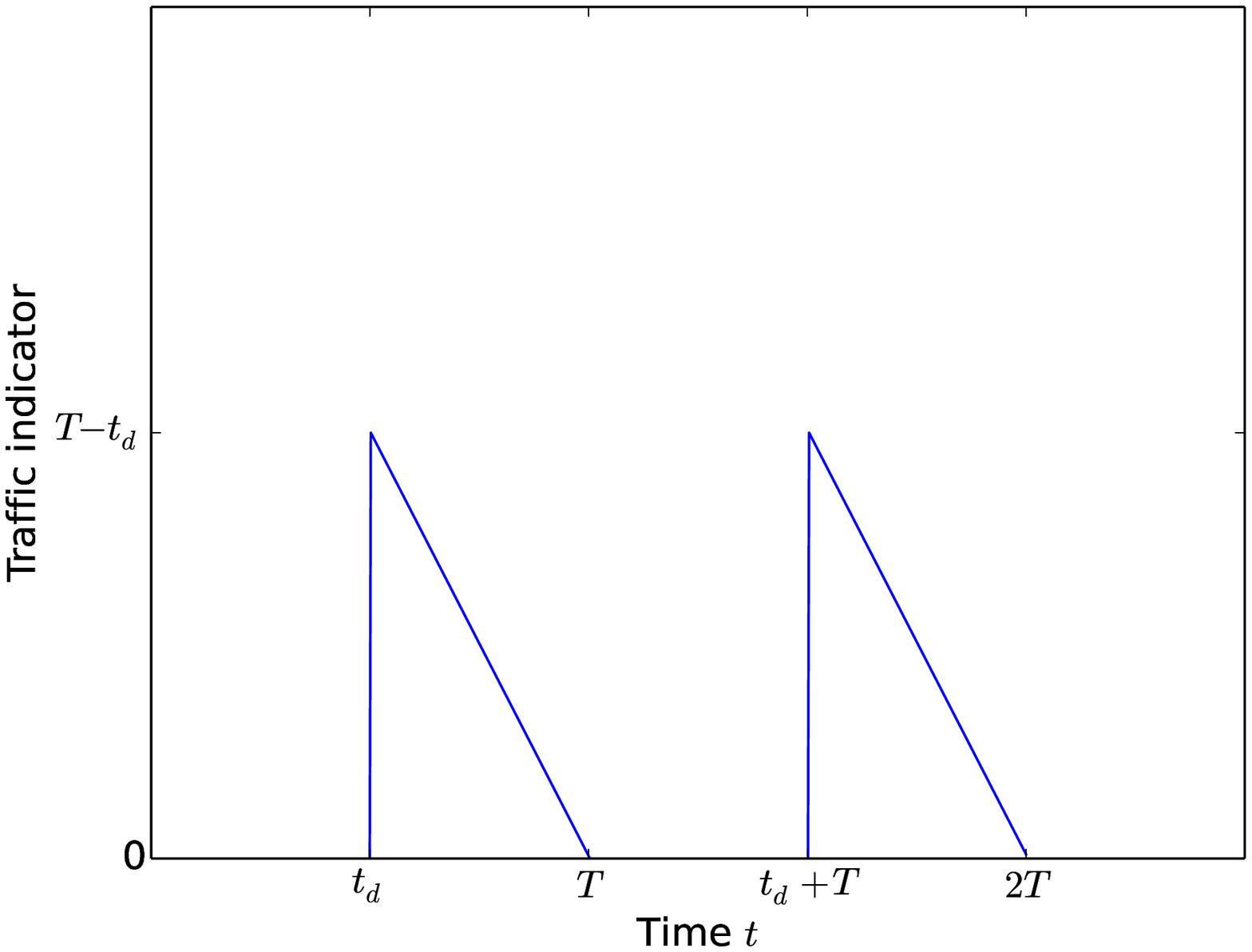}
}
  \end{minipage}%
\hfill
  \begin{minipage}[t]{0.33\linewidth}
  \centering
\subfigure[flow $f_2(t)$]{
\label{fig:flow_one}
\includegraphics[width=1\linewidth]{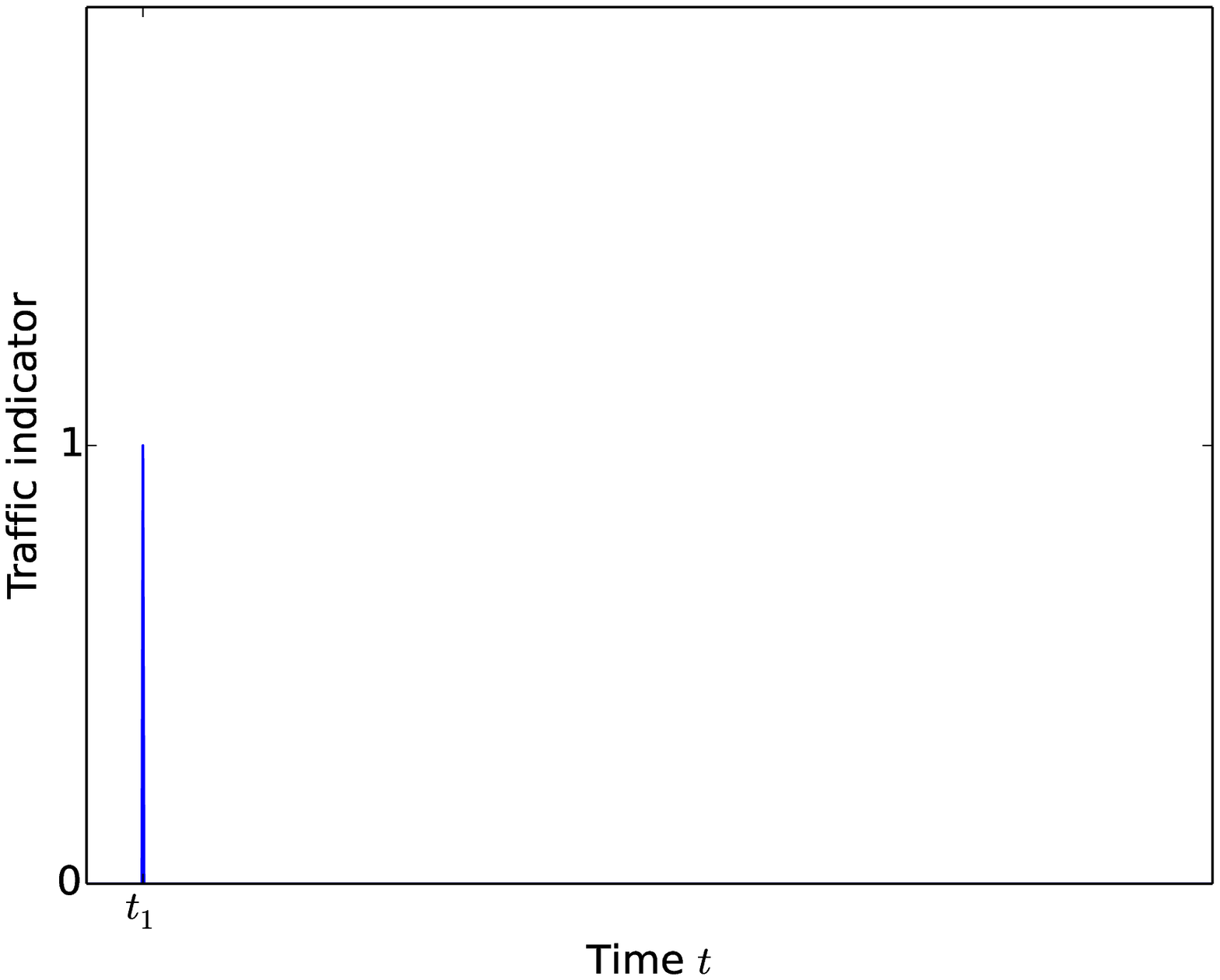}
}
  \end{minipage}
\hfill
    \begin{minipage}[t]{0.33\linewidth}
    \centering
  \subfigure[flow $f_3(t)$]{
  \label{fig:flow_three}
  \includegraphics[width=1\linewidth]{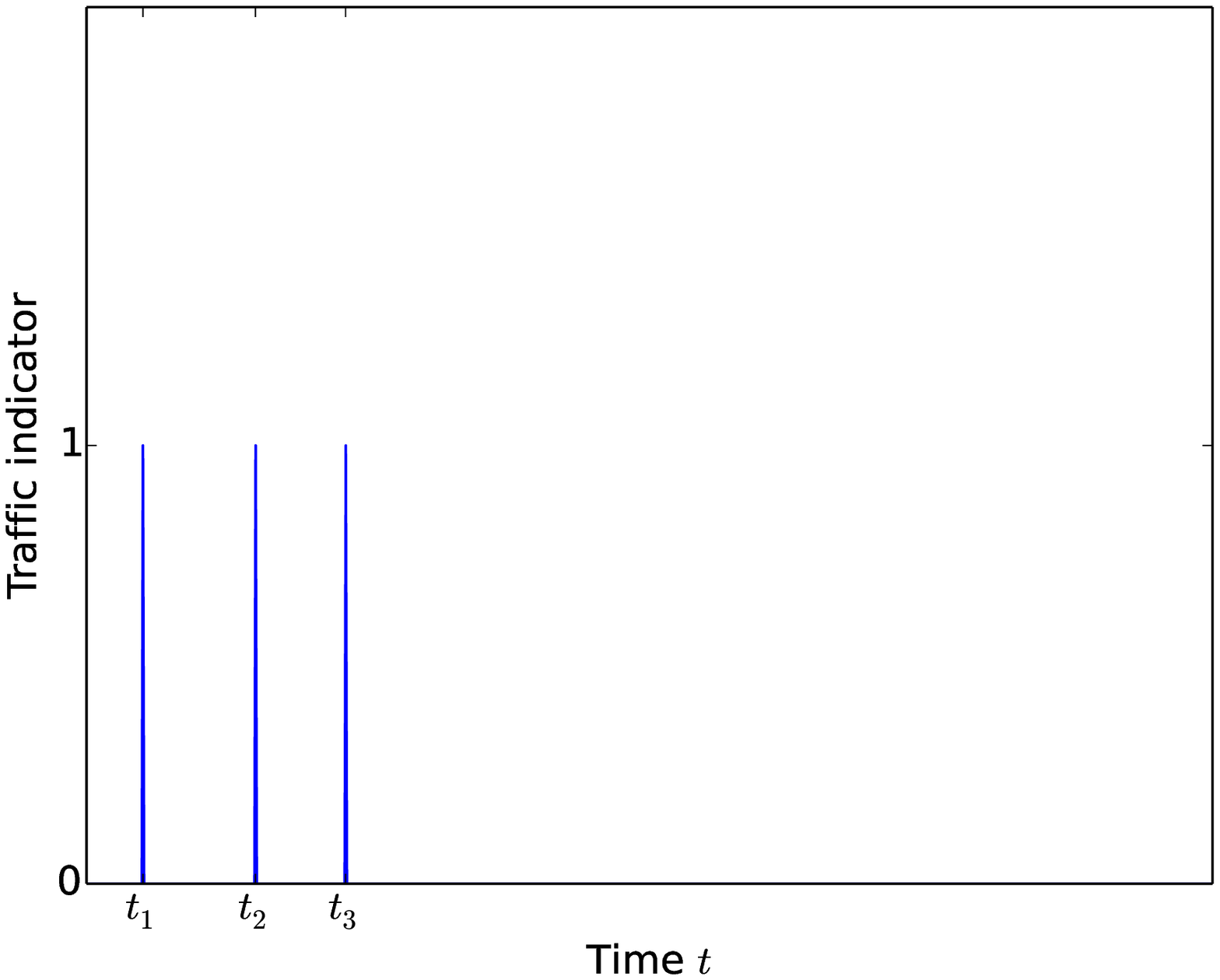}
  }
    \end{minipage}
    \begin{minipage}[t]{0.33\linewidth}
    \centering
  \subfigure[$T(f_1, t)$]{
  \label{fig:func_predict}
  \includegraphics[width=1\linewidth]{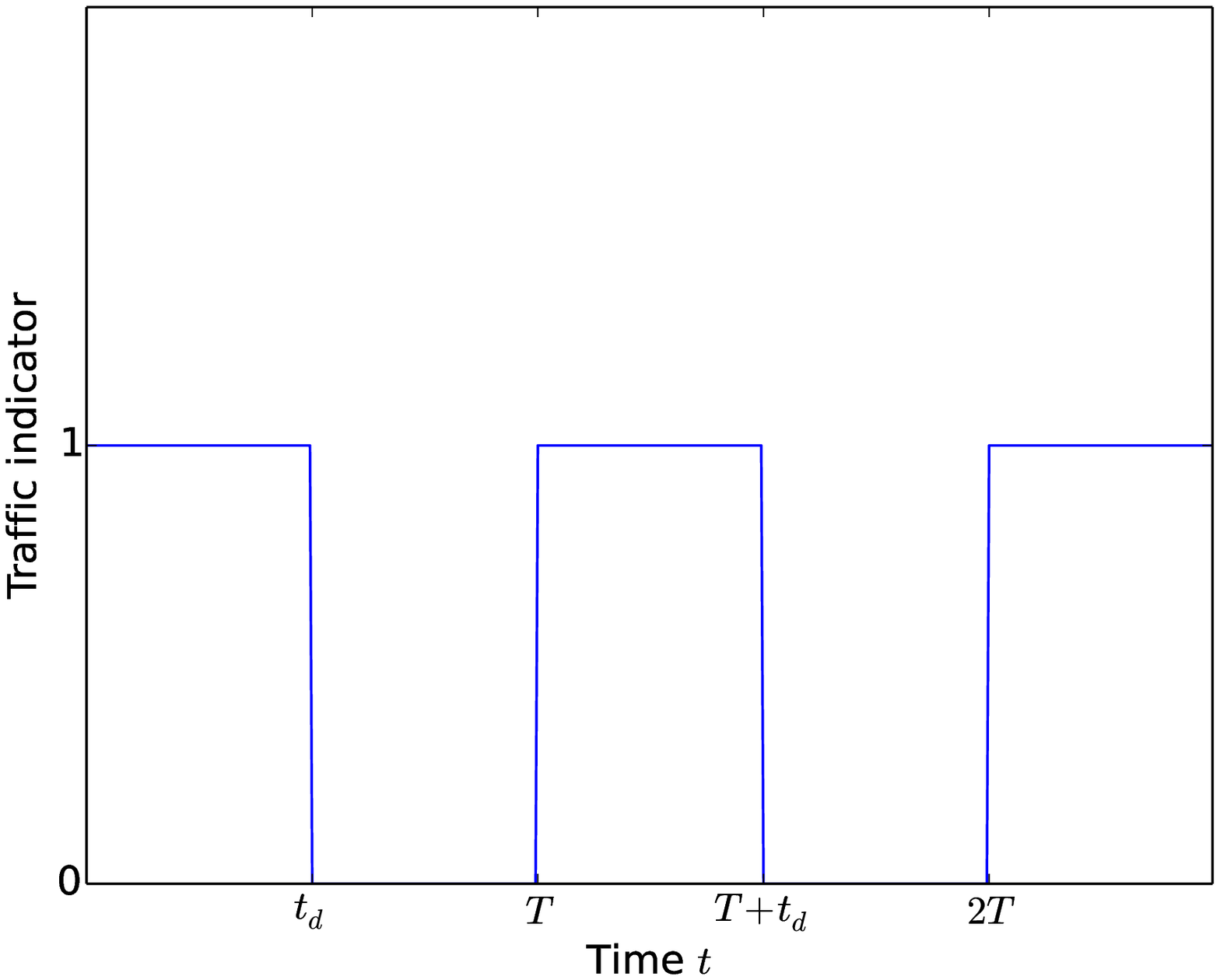}
  }
    \end{minipage}%
    \hfill
    \begin{minipage}[t]{0.33\linewidth}
    \centering
  \subfigure[$T(f_2, t)$]{
  \label{fig:func_onetime}
  \includegraphics[width=1\linewidth]{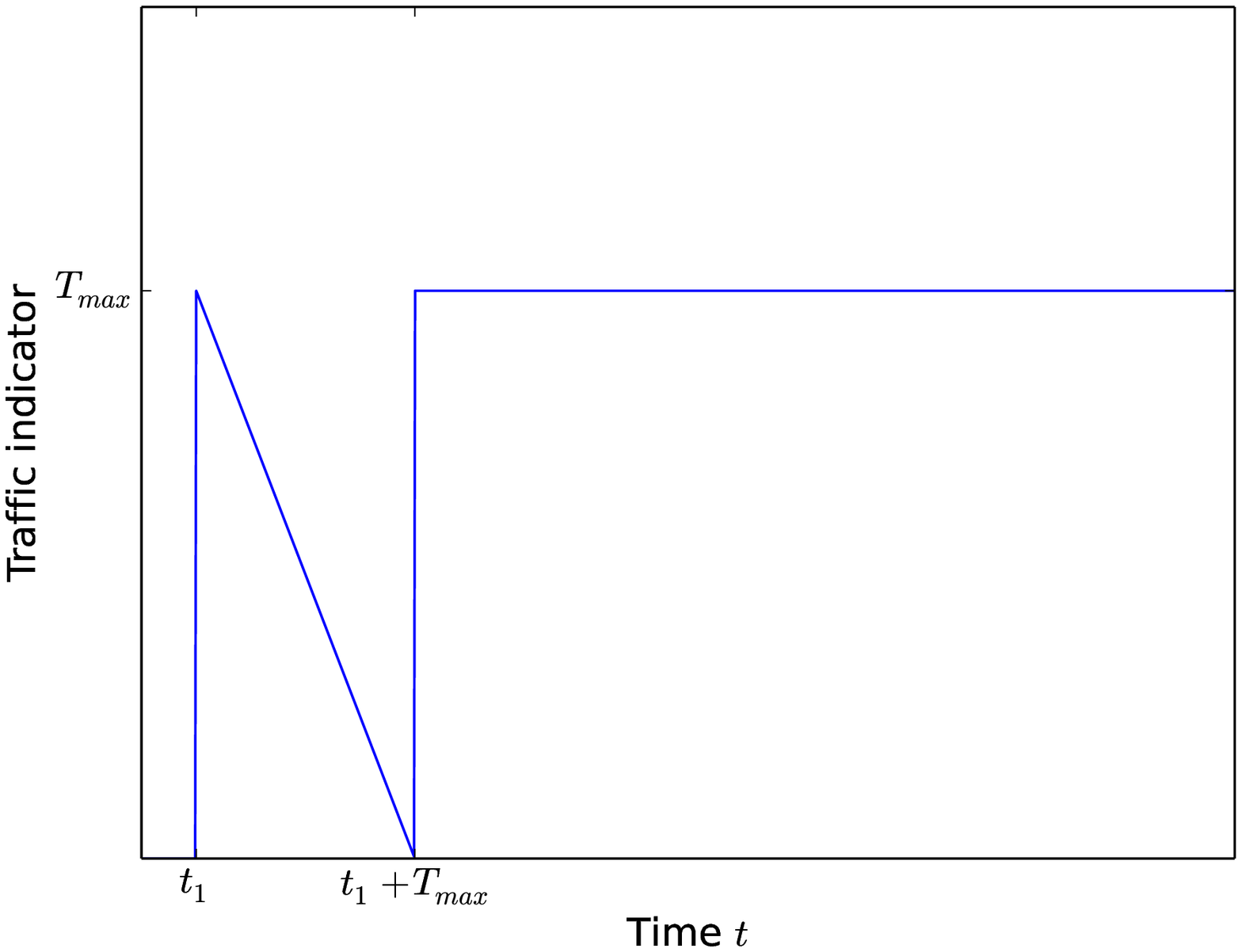}
  }
    \end{minipage}
      \hfill
      \begin{minipage}[t]{0.33\linewidth}
      \centering
    \subfigure[$T(f_3, t)$]{
    \label{fig:func_unpredict}
    \includegraphics[width=1\linewidth]{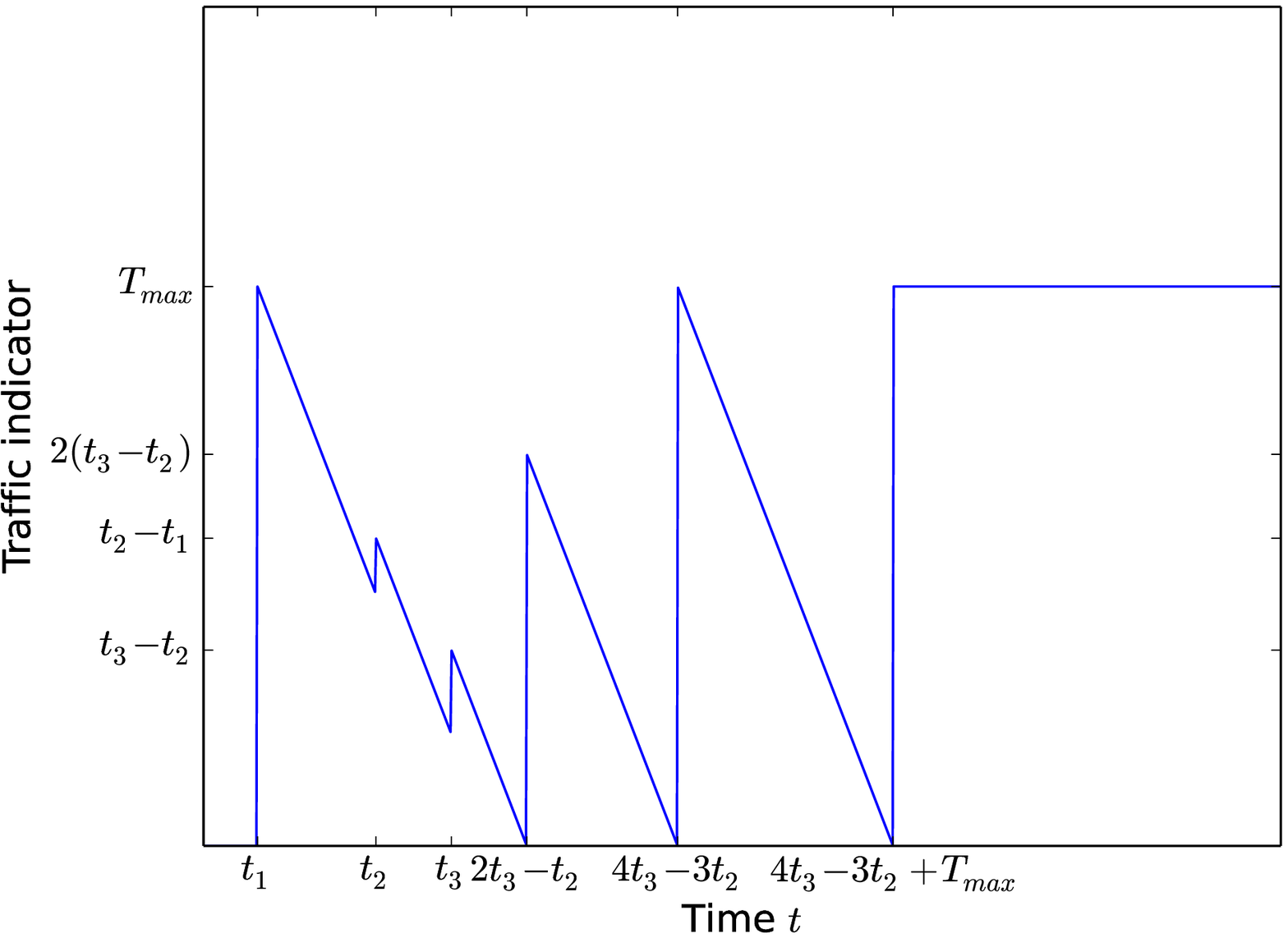}
    }
      \end{minipage}

\centering

  \caption{Examples of $T(f_i, t)$ for various types of flows}
  \label{fig:time_func}
\end{figure*}

In this section, we propose a \textit{Flow-Driven Caching} (FDRC) algorithm to solve the rule caching problem by taking into account of both traffic pattern and routing path of each flow. The basic idea of our algorithm is as follows. When the rule for any flow $f_i$ needs to be cached on switch $s_j$, a timer $T_i$ is associated to the entry with a value that is an estimated time to the next hit of the entry. (2) When an entry replacement happens at switch $s_j$, the entry with maximum timer value will be chosen. The description  of FDRC is given in Algorithm \ref{alg:caching}.

\newcommand{\argmax}{\mathop{\rm arg~max}\limits}
\newcommand{\argmin}{\mathop{\rm arg~min}\limits}

\begin{algorithm}[t]
\centering
\caption{Flow-driven caching algorithm}
\label{alg:caching}
\begin{algorithmic}[1]
\Require{In time $t$, a packet of flow $f_i$ comes to the switch.}
\For {$s_j \in S_i}$ \Comment{Traverse switches in the forwarding path }
\label{alg_line:traverse} 
	\If {$r_i$ is not (i.e., $X_{ij}=0$) cached in $s_j$}  
		\If{$\sum_{i=1}^{n}X_{ij} = B_j$}\Comment{Cache replacement}
		\label{alg_line:replace}
			
			\State  $m=\argmax_{\ \ \ k\ \ X_{kj}=1}T_k$;
		
			\State remove $r_m$ from cache;
		
		\EndIf
	\State put $r_i$ in cache;
    \State Set timer $T_i$ using Algorithm \ref{alg:cach}
	\label{alg_line:put_finish}
	\EndIf
	\State forward the packet using $r_i$
\EndFor
\end{algorithmic}

\end{algorithm}

When setting timer $T_i$, we consider two type of flow patterns: predictable flows (e.g., the flows from deterministic network service) and unpredictable flows (e.g., spontaneous traffic). For the former case, the timer is simply set as the time to netxt packet arrival. For the letter case, the estimation of the time is updated using Algorithm \ref{alg:cach}.
\begin{algorithm}[t]
\centering
\caption{Setting times $T_i$ for unpredicted flow $f_i$}
\label{alg:cach}
\begin{algorithmic}[1]
\State Start time $T_i$ with an initialized value $T_max$;
\While {1}
\If {a new packet of flow $f_i$ comes before $T_i$ expires} 
\State set $\Delta T$ as the latest arriving packet interval; 
\State start $T_i$ with value $\Delta T$;
\EndIf
\If {$T_i$ expires} 
\If {$\Delta T = T_{max}$}
\State freeze $T_i$ with value $T_{max}$;
\State break;
\EndIf
\State $T_i$ with value $\min(2\Delta T, T_{max})$;
\EndIf
\EndWhile
\end{algorithmic}
\end{algorithm}

From Algorithm \ref{alg:cach}, we notice that when the estimated next arrival is earlier than expected, the times should be updated by the latest arriving interval; otherwise, the timer should be reset with a double value, but no exceeding $T_{max}$, after expiration.

To better understanding how $T_i$ behavior, we show its value are function of time t, denoted as $T(f_i, t)$, using example in Figure \ref{fig:time_func}. In Figure  \ref{fig:time_func}\subref{fig:flow_predict}-\subref{fig:flow_three}, the values 1/0 on y-axis indicate if a flow is active or not at any time t. The corresponding functions $T(f_i, t)$ are illustrated in Figure \ref{fig:time_func}\subref{fig:func_predict}-\subref{fig:func_unpredict}, respectively.

Figure \ref{fig:flow_one} shows a predictable flow $f_1$, which keeps active for a period $t_d$ and then silent for $T-t_d$, periodically. Therefore, $T(f_1, t)$ in Figure \ref{fig:func_predict} is also a periodic function, with a cycle length $T$, that can be determined as follows.
\begin{eqnarray}
 & T(f_1, t) =
\begin{cases}
 0 \    t \in (nT, nT+t_d)\\
 (n+1)T+t_d - t \ t \in (nT+t_d, (n+1)T)
\end{cases}\nonumber\\
& n=0, 1, 2, ... \nonumber
\end{eqnarray}

Example of unpredictable flows are given in Fig. \ref{fig:time_func}\subref{fig:flow_one} and \subref{fig:flow_three}. For example, when a packet of $f_2$ arrives at $t_i$ shown in Figure \ref{fig:flow_one}, time $T_2$ is set as $T_{max}$, i.e., 
\[
T(f_2, t) = T_{max} + t_1 - t\text{.}
\]
According to Algorithm 2, after $T_2$ expires, $T(f_2, t)$ remains $T_{max}$ as shown in Figure \ref{fig:func_onetime}. On the other hand, when a packet of $f_3$ arrives at $t_2$ earlier than the estimated time, $T(f_3, t)$ should be updated from $T(f_3, t) = T_{max} + t_1 -t$ to $T(f_3, t) = (t_2 - t_1) + t_2 - t$ as shown in Figure \ref{fig:func_unpredict}. Later on, $T_3$ will expire at $2t_3-t_2$ and the restart from  $\min(2(t_3-t_2), T_{max}) = 2(t_3-t_2)$. The second expiration happens at $4t_3-3t_2$ and $T_3$ restarts from $\min(4(t_3-t_2)), T_{max}) \geq T_{max}$. Eventually, it remains at $T_{max}$.

\section{Evaluation}
\label{sec:s}
We conduct simulation based experiments to evaluate the performance of the proposed algorithm. Simulation results under different network parameters are presented.

\subsection{Simulation Settings}

To evaluate the performance of our algorithm, we compare the cache-hit of different caching algorithm over a number of randomly generated networks by using a python 2.7 script with the network library 1.6 on a desktop computer. We use two types of flows in the simulation, flows with periodic traffic and random traffic. For the periodic traffic, the period $T$ of the traffic cycles are uniformly distributed within the range $[2s, 100s]$ and traffic duration time in each cycle is uniformly distributed within range $[1s, T]$. For the random traffic, the packets are generated randomly and the interval periods between packets are uniformly distributed in range from 0 to the simulation end time.

The number of related switches for each flow is normally distributed in range $[1, 10]$. The simulation generates these flows randomly and puts them to the network. The default simulation setting is as follows.

(1) result from the first hour traffic 

(2) cache size of the SDN switches is normally distributed in range $[15,25]$

(3) ratio of predictable flows in total flows is 40\%.

For the purpose of comparison, the following algorithms are also considered:

(1) FIFO, which is the default algorithm in SDN switches.

(2) LRU, a popular caching algorithm. 

All simulation results are averaged over 20 network instances.

\subsection{Simulation Results}

\begin{figure}[t]
\centering
\includegraphics[width=0.7\linewidth]{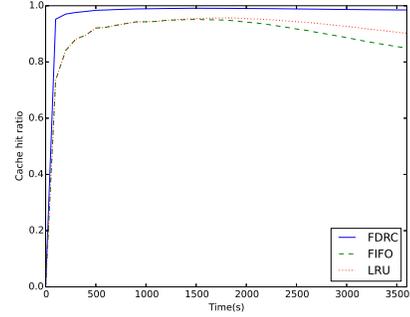}
\caption{Cache hit ratio in the first hour}
\label{fig:cache_ratio_time}
\end{figure}

First, we test the performance with default setting. As shown in Figure \ref{fig:cache_ratio_time}, the result of cache hit ratio with the default setting shows our FDRC algorithm has better performance than the other two algorithms, in terms of both maximum ratio and how fast this ratio can be archived. After 1500s, the hit ratio with FDRC is also better than the ratio with FIFO and LRU. FIFO brings lowest hit ratio in these three algorithm, and LRU performs 6\% better than FIFO.

\begin{figure}[t]
\centering
\includegraphics[width=0.7\linewidth]{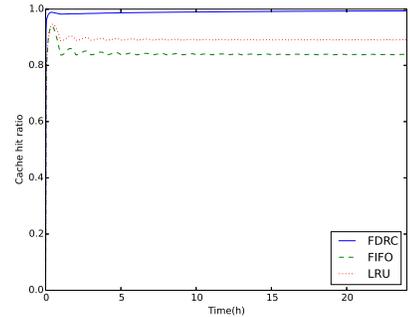}
\caption{Cache hit ratio per one day period}
\label{fig:cache_hit_day}
\end{figure}

Then, we extend the flow traffic time to test the performance in a longer period shown in Figure \ref{fig:cache_hit_day}. At the beginning of the day, new flows are put to the network and to full fill the cache of switches. When the cache of each switch become full, the hit ratio decreases with incorrect replacements. At last, the ratio become stable with the balance between correct and incorrect replacements. In the final of the day, the hit ratio with FDRC is near to 99.4\% while the ratio with FIFO and LRU is near to 83.8\% and 89.0\%.

\begin{figure}[t]
\centering
\includegraphics[width=0.7\linewidth]{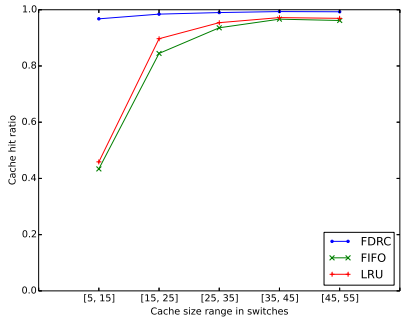}
\caption{Cache hit ratio with different cache size in switches}
\label{fig:cache_hit_size}
\end{figure}

Third, since the cache size has a positive relationship with the cache hit ratio, we test the cache ratio with different cache size. as shown in Figure \ref{fig:cache_hit_size}. We use five ranges of the switch cache size which are $[5, 15]$, $[15, 25]$, $[25, 35]$, $[35, 45]$ and $[45, 55]$. The result shows the FDRC algorithm brings better cache hit ratio with small cache size than other two algorithms. When the cache size is very small in range $[5, 15]$, the cache hit ratio with FIFO and LRU is lower than 5\%, which means the cache size has  effect on the network performance. Therefore, with a small cache size, the hit ratio with FDRC algorithm is larger than 90\%. With an increased cache size, the differences among these algorithms become small. With range $[45, 55]$ of the cache size, FDRC brings about 3.2\% better than the other two algorithms.

\begin{figure}[t]
\centering
\includegraphics[width=0.7\linewidth]{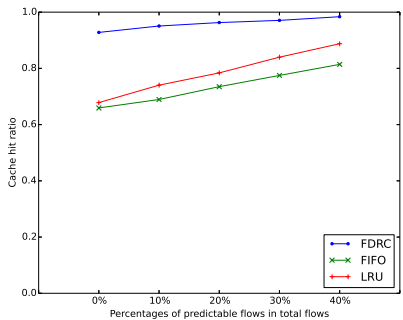}
\caption{Cache hit ratio with different percentages of predictable flows in all flows}
\label{fig:cache_hit_percentage}
\end{figure}

Last, since our algorithm adopts different strategies for predictable and unpredictable flows, we adjust the ratio of predictable flows in total flows and test the cache hit ratio. We use five counts of the percentage of predictable flows in all flows and test the cache hit ratio with these percentages. From the result shown in Figure \ref{fig:cache_hit_percentage}, the cache hit ratio with FDRC is better than the ratio with other two algorithm especially with unpredictable flows. When no predictable flow is given in the network, FDRC brings better performance than other algorithm. When the percentage of predictable flows increases, the hit ratio becomes higher with these three algorithm. When the percentage of predictable becomes 40\%, the cache hit ratio with FDRC is still 10\% and 15\% better than the ratio with LRU and FIFO, respectively.

\subsection{Analysis}

From the evaluation on the cache hit ratio, we find our algorithm brings better performance than other two algorithm especially in longer period with better processing on the predictable and unpredictable flows. The cache size also influent seriously to the hit ratio with the traditional cache algorithms since they are short of any prefetching optimization on the path of each flow. With the percentage of the predictable flows increases, three algorithm brings better performance especially the LRU since the predictable flows appear frequently in an shorter period than unpredictable flows with random traffic distributions.

\section{Conclusions}
\label{sec:c}

In this paper, we propose a rule caching model based on the traffic and path of flows to optimize replacement of a switch cache. We apply the prefetching with the path of each flow to reduce that cache miss during forwarding this flow in its path. To meet the predictability of flows in SDN structure, we also design some special processing on the predictable and unpredictable flows. We study a rule caching problem to maximize the cache hit ratio of the SDN-based network. Finally, extensive simulations are conducted to show that the proposed caching algorithm can significantly increase the hit ratio than traditional algorithms.

\section*{Acknowledgement}
This work is partially sponsored by the National Basic Research 973 Program of China (No. 2015CB352403), the National Natural Science Foundation of China (NSFC) (No. 61261160502, No. 61272099, No. 61303012, No.61332001), the Program for Changjiang Scholars and Innovative Research Team in University (IRT1158, PCSIRT), the Scientific Innovation Act of STCSM (No. 13511504200), the Shanghai Natural Science Foundation (No. 13ZR1421900), the Scientific Research Foundation for the Returned Overseas Chinese Scholars, and the EU FP7 CLIMBER project (No. PIRSES-GA-2012-318939)

\bibliographystyle{IEEEtran}
\bibliography{rule_caching}

\end{document}